\documentclass[prl,twocolumn,floatfix,superscriptaddress]{revtex4-2}

\usepackage{amsmath, amssymb}
\usepackage{enumitem}
\usepackage{mathtools}
\usepackage{lipsum}
\usepackage{times}
\usepackage{graphicx}
\usepackage{wasysym}
\usepackage{bm}
\usepackage[hidelinks]{hyperref}
\usepackage{xspace}
\usepackage{siunitx}
\usepackage{xcolor}
\usepackage{soul}
\usepackage{braket}
\usepackage[normalem]{ulem}
\usepackage[T1]{fontenc}

\definecolor{myColor}{rgb}{0,0,0.8}%{0.02,0.12,0.3}
\definecolor{myciteColor}{rgb}{0.39,0.7,0.89}
\hypersetup{colorlinks=true, linkcolor=myColor, filecolor=myColor, urlcolor=myColor,citecolor=myColor, urlcolor=myColor}
\graphicspath{{./}}

\DeclareSIUnit{\nK}{\nano\kelvin}
\DeclareSIUnit{\aB}{\emph{a}_0}
\DeclareSIUnit{\G}{G}

\renewcommand{\figurename}[1]{Fig.~}

\newcommand{\Nc}{N_\mathrm{c}}

\DeclareSIUnit\litre{l}
\newcommand{\IAP}{Institut für Angewandte Physik, Universität Bonn, Wegelerstrasse 8, 53115 Bonn, Germany}

\begin{document}

\title{Bose-Einstein Condensation of Photons in a Four-Site Quantum Ring}

\author{Andreas Redmann}
\email[]{redmann@iap.uni-bonn.de}
\affiliation{\IAP}
\author{Christian Kurtscheid}
\altaffiliation[Present address: ]{Fraunhofer-Institut für Hochfrequenzphysik und Radartechnik FHR, Fraunhoferstr. 20, 53343 Wachtberg, Germany}
\affiliation{\IAP}
\author{Niels Wolf}
\affiliation{\IAP}
\author{Frank Vewinger}
\affiliation{\IAP}
\author{Julian Schmitt}
\affiliation{\IAP}
\author{Martin Weitz}
\affiliation{\IAP}

\date{\today}

\begin{abstract}
Thermalization of radiation by contact to matter is a well-known concept, but the application of thermodynamic methods to complex quantum states of light remains a challenge. Here we observe Bose-Einstein condensation of photons into the hybridized ground state of a coupled four-site ring potential. In our experiment, the periodically-closed ring lattice superimposed by a weak harmonic trap for photons is realized inside a spatially structured dye-filled microcavity. Photons thermalize to room temperature, and above a critical photon number macroscopically occupy the symmetric linear combination of the site eigenstates with zero phase winding, which constitutes the ground state of the system. The mutual phase coherence of photons at different lattice sites is verified by optical interferometry.
\end{abstract}
%
%%\keywords{}
%

\maketitle

%\js{\cs\cs INTRO1: MACROSCOPIC GROUND STATES}
Populating the ground state of complex systems by freezing out thermal excitations is at the core of many low-temperature quantum phenomena that involve correlated many-body states, such as fractional quantum Hall or spin-charge separated states~\cite{Chang:2003, Deshpande:2010}. A tailoring of correlated quantum states has also been achieved with cold atoms in optical lattice potentials, \emph{e.g.}, to realize effects known from solid state physics as the Mott-insulator transition~\cite{Greiner:2002a}. An accumulation into the ground state in the periodic potential is here achieved by adiabatic loading of cold atoms into the lattice potential.

%\js{\cs\cs INTRO2: STATE-OF-THE-ART IN OPTICAL QUANTUM GASES}

A different route to study many-body physics by direct cooling has recently been explored in the field of optical physics. Studies in this domain have long been restricted to nonequilibrium phenomena as coupled laser systems~\cite{Bao:2004,Nixon:2013}, but more recent work has realized thermalization of low-dimensional optical quantum gases, \emph{e.g.} in exciton-polariton gases by interparticle collisions~\cite{Deng:2010,Carusotto:2013} and in photon gases by contact to an equilibrium reservoir~\cite{Klaers:2010,Schmitt:2018,Bloch:2022}, opening new ways of cooling directly into nontrivial many-body states. The latter have been demonstrated to allow for a thermodynamic phase transition of photons to a macroscopically occupied ground state, the Bose-Einstein condensate. Corresponding experiments have been carried out in dye-filled optical microcavities~\cite{Klaers:2010,Marelic:2015,Greveling:2018,Vretenar:2021a}, erbium-doped fibers~\cite{Weill:2019}, plasmonic lattices ~\cite{Hakala:2018}, and more recently in semiconductor microcavities~\cite{Schofield:2024,Pieczarka:2024}. Furthermore, theoretical work has suggested Bose-Einstein condensation of photons in a microcavity plasma~\cite{Figueiredo:2023}. Such light-matter platforms open the possibility to engineer quantum states of light by energetic selection~\cite{Klaers:2012}, while state preparation methods based on the competition between gain and loss are limited to classical physics phenomena~\cite{Berloff:2017,Gershenzon:2020}. In a recent experiment, the thermalization of light in a double well potential to populate the “bonding” ground state of this optical analog of a dimer molecule has been demonstrated~\cite{Kurtscheid:2019}. On the way to more complex systems, lattice potentials have been realized with both photon and polariton systems~\cite{Jacqmin:2014,Vretenar:2021b,Lagoudakis:2017}. With polaritons, both thermalization to the bottom edge of a tilted ring structure as well as multi-mode condensation into coupled states of opposite orbital flow in a ring have been investigated~\cite{Mukherjee:2019a, Mukherjee:2021, Wang:2021a}.

%\js{\cs\cs SHORT RATIONALE}

In this Letter, we report measurements of the thermalization and Bose-Einstein condensation of a two-dimensional photon gas to the ground state of a periodically-closed ring lattice, where the notation of a phase winding becomes meaningful. The four-site ring is superimposed by a harmonic trapping potential and is realized by imprinting a corresponding surface structure on one of the cavity mirrors of a dye-filled optical microcavity. The photon gas thermalizes to room temperature by repeated absorption and emission on the dye molecules. Above a critical photon number, we observe a macroscopic fraction of photons occupying the linear combination of eigenfunctions with zero phase winding, which constitutes the energetic ground state of the system. This is accompanied by a saturation of the population in the excited modes, as expected for the phase transition to a Bose-Einstein condensate. The photons delocalize over the ring structure and the obtained fixed phase relation between photons from different lattice sites is verified by optical interferometry.

%\js{\cs\cs EXPERIMENTAL SETTING}

\begin{figure}[t]
    \centering
    \includegraphics[width=1.0\columnwidth]{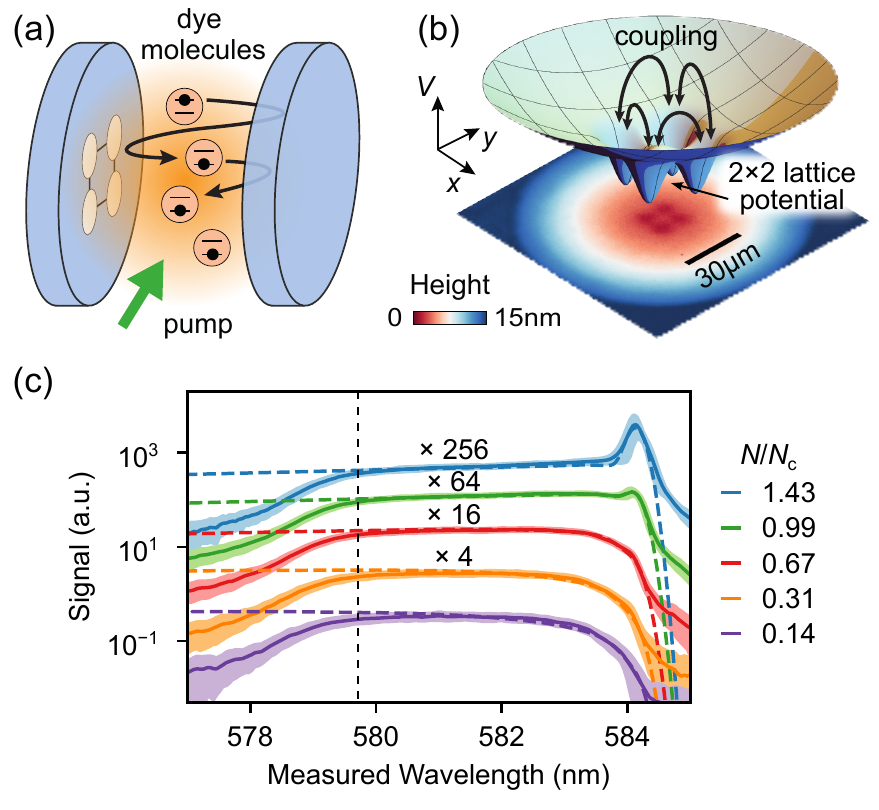}
\caption{(a)~Experimental scheme of dye-filled optical microcavity formed by a microstructured and a plane mirror. The confined photon gas thermalizes by absorption and re-emission processes on dye molecules. (b)~Potential experienced by the photons in the cavity (top), as induced by the structured mirror surface (bottom). The coupling between neighboring sites realizes a four-site ring lattice with hybridized eigenstates. (c)~Spectral distributions of the thermalized photon gas for increasing $N/\Nc$ along with theoretical Bose-Einstein spectra (dashed lines) which take into account the measured mode densities and spectrometer resolution. The finite depth of the imprinted potential is indicated by the black dashed line. Standard deviations are shown as shading.}
    \label{fig:1}
\end{figure}

The used dye microcavity platform shown in Fig.~\ref{fig:1}(a) confines photons between two highly reflecting mirrors of reflectivity above 99.997\%, where the cavity volume is filled with a Rhodamine 6G dye solution of \SI{1}{\milli\mole\per\litre} concentration in ethylene glycol (refractive index $\tilde n\approx 1.44$). The microcavity length $D_0 \approx \SI{1.8}{\micro\meter}$ is in the wavelength regime, which results in a longitudinal mode spacing sufficiently large that only a single longitudinal mode, with mode number $q=9$, is populated. This makes the photon gas effectively two-dimensional, introduces a low-frequency cutoff near $\hbar \omega_\mathrm{c} \approx \SI{2.1}{\electronvolt}$, with Planck's reduced constant $\hbar$, and the dispersion becomes quadratic. To implement a four-site ring lattice potential superimposed with a harmonic trap, one of the two cavity mirrors is microstructured. For this, a position-dependent local surface elevation is imprinted on the reflecting mirror surface, see Fig.~\ref{fig:1}(b). Inside the assembled microcavity the elevation leads to a locally repulsive potential, as understood from the shorter wavelength (higher frequency) of the photons required to match the cavity boundary conditions at the corresponding elevated transverse position~\cite{Kurtscheid:2019,Kurtscheid:2020}. Accounting for the boundary condition for the longitudinal wavevector component $k_z = q \pi / \{\tilde{n} [D_0 - d(x,y)]\}$ imposed by the mirrors, where $d(x,y)$ denotes the position-dependent mirror elevation and $d(x,y) \ll D_0 $, the eigensolutions of photons in the resonator are in paraxial approximation determined by the effective two-dimensional Hamiltonian~\cite{Schmitt:2018,Kurtscheid:2019}
\begin{align}
	\hat{H}_{x,y} = m_\mathrm{ph} \left(\frac{c}{\tilde{n}}\right)^2 - \frac{\hbar^2}{2 m_\mathrm{ph}} \left(\nabla_x^2 + \nabla_y^2\right) + V(x,y) \, ,
\end{align}
noting that interactions are weak. Here $c$ denotes the speed of light, $m_\mathrm{ph} = \pi \hbar q \tilde{n} / (c D_0) = \hbar \omega_\mathrm{c} (\tilde{n}/c)^2$ an effective photon mass, and $V(x,y) = m_\mathrm{ph} {(c/\tilde{n})^2} d(x,y)/D_0$ the imprinted trapping potential. Correspondingly, the photon gas in the cavity is formally equivalent to a two-dimensional gas of massive bosons subject to the potential $V(x,y)$. Thermalization of photons to the rovibrational temperature of the dye~\cite{Klaers:2012}, which is at room temperature, is achieved by repeated absorption re-emission processes on the dye molecules provided the photon lifetime in the cavity is sufficient~\cite{Kirton:2015,Schmitt:2015}. To inject an initial photon population and compensate for photon loss from, \emph{e.g.}, mirror transmission, the dye microcavity is quasi-continuously pumped with a laser beam at $\SI{532}{\nano\meter}$ wavelength and $\SI{55}{\micro\meter}$ diameter (FWHM).

%%%%%%%%%%%%%%%%%
%%%%% FIG 1 %%%%%
%%%%%%%%%%%%%%%%%

%\js{\cs\cs THEORY OF THE RING}

We model the four-site system using an ansatz for the (scalar) wavefunction of photon eigenstates $\psi(x,y) = \sum_i c_i \varphi^{(i)} (x,y)$, where the $\varphi^{(i)}(x,y) = \varphi(x - x_i,y - y_i)$ denote localized wavefunctions and the $c_i$ probability amplitudes, with the index $i$ numbering the sites. The coupling between sites leads to a hybridization of the localized wavefunctions. The eigensolutions of photons in the lattice can straightforwardly be evaluated by diagonalization of the effective Hamiltonian
\begin{align}
	\hat H_\mathrm{lat} \simeq 
    \begin{pmatrix}
		E_\mathrm{s} & -J & 0 & -J \\
		-J & E_\mathrm{s} & -J & 0 \\
		0 & -J & E_\mathrm{s} & -J \\
		-J & 0 & -J & E_\mathrm{s}
	\end{pmatrix} \, .
	\label{eq:Hamiltonian}
\end{align}
The four-component Hamiltonian acts on spinors $\psi = (c_1, c_2, c_3, c_4)$. Here $E_\mathrm{s}$ denotes the localized wavefunctions eigenenergy, and we only account for the coupling $J$ between neighboring sites, with $J = \bra{\varphi^{(i)}} \hat{H} \ket{\varphi^{(i \pm 1)}}$, and use a cyclic notation ($\varphi^{(1)} \equiv \varphi^{(5)}$). A set of eigenstates, each of them possessing a two-fold polarization degeneracy, can now be written in the form $\psi_{\ell} = \frac{1}{2} \left(\varphi^{(1)} + e^{i \frac{\pi}{2}\ell} \varphi^{(2)}  + e^{i \pi \ell} \varphi^{(3)}  + e^{i \frac{3\pi}{2}\ell } \varphi^{(4)} \right)$ that is ordered by the phase winding of the four-site quantum ring with $\ell=0,\pm 1,2$. Specifically, for $\ell = 0$, the system ground state at energy $E_\mathrm{s} - 2J \left(\equiv \hbar \omega_\mathrm{c} \right)$ is the symmetric linear combination $\psi_0 = \frac{1}{2} \left(\varphi^{(1)} + \varphi^{(2)}  + \varphi^{(3)} + \varphi^{(4)} \right)$, and the excited states of the lattice $\psi_{\pm 1}$ and $\psi_{2}$ have the energies $E_\mathrm{s}$ and $E_\mathrm{s} + 2J$, respectively.

Experimentally, the ring potential for the photons described by eq.~\eqref{eq:Hamiltonian} is realized by imprinting a surface profile with four indents spaced by roughly \SI{10}{\micro\meter} onto one of the cavity mirrors~\cite{Kurtscheid:2020}, see Fig.~\ref{fig:1}(b). The indents each exhibit a width of \SI{6.7}{\micro\meter} (FWHM) and a depth of \SI{0.8}{\nano\meter}, which yields a ground state energy close to the potential energy of the barrier separating nearest-neighbor lattice sites but below the potential barrier located in the trap center~\cite{Supplementary}. The coupling between neighboring sites, as determined spectroscopically, reaches $J \approx h \cdot \SI{25}{\giga\hertz}$. The lattice potential is superimposed by a harmonic trap with angular frequency $\Omega = c / (\tilde{n}\sqrt{D_0 R}) \approx 2 \pi \cdot \SI{64}{\giga\hertz}$, as realized by additionally imprinting a spherical curvature with a radius of of curvature of $R \approx \SI{14.8}{\centi\meter}$ on top of the lattice indent structure. The mode density of the harmonic trap is required to enables Bose-Einstein condensation of photons in the two-dimensional system~\cite{Kurtscheid:2019}.

%\js{\cs\cs EXPERIMENT 1: THERMALIZED SPECTRA}

\begin{figure}[t]
    \centering
    \includegraphics[width=1.0\columnwidth]{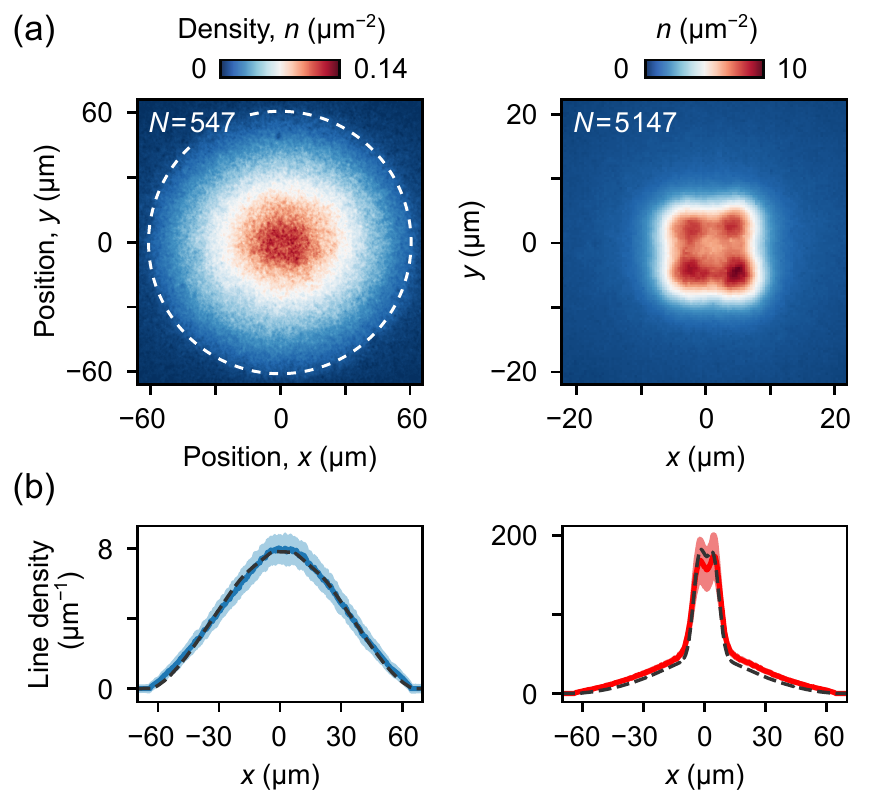}
\caption{(a)~Density distribution of the photon gas in the microcavity both below (left panel) and above (right) the Bose-Einstein condensation threshold. For $N/\Nc = \SI{0.20 \pm 0.02}{}$, one observes a rotationally symmetric thermal distribution, while for $N/\Nc = \SI{1.8 \pm 0.3}{}$, the four-fold symmetry of the lattice becomes visible as attributed to the dominant contribution by the symmetric ground state $|\psi_0|^2$. Dashed circle indicates the radial extent of the finite-depth harmonic trap. (b)~Line densities obtained from integration along the vertical axis of the distributions shown in (a), along with fits based on room-temperature Bose-Einstein distributions (dashed lines), yielding $N=\SI{510 \pm 40}{}$ and $N=\SI{4200 \pm 200}{}$ as fit parameters. Standard deviations are shown as shading.}
    \label{fig:3}
\end{figure}

We first verify the thermalized nature of the photon gas in the combined harmonic and four-site ring potential, being a prerequisite for Bose-Einstein condensation. For this, we have recorded broadband spectra (spectrometer resolution $\SI{300}{\giga\hertz}$) for different photon numbers, see Fig.~\ref{fig:1}(c). The spectrometer allows us to record the spectral distribution of the photon gas over the entire trap depth of the harmonic potential. The data is in good agreement with the theory expectation for a Bose-Einstein distribution at $T=\SI{300}{\kelvin}$ over the spectral range from $\SI{584.1}{\nano\meter}$ (corresponding to the low-frequency cutoff) to $\SI{579.7}{\nano\meter}$, at which the finite trap depth of the imprinted potential of approximately $0.63 \cdot k_\mathrm{B}T\approx h\cdot \SI{3.9}{\tera\hertz}$ is reached. The data provides a first line of evidence for the thermalization of photons in the tailored potential, which will be complemented by measurements of the spatial intensity distributions in the following, see Fig.~\ref{fig:3}. While at low photon numbers a Boltzmann distribution is observed, for photon numbers exceeding a critical number $\Nc \approx 2780$ a macroscopic occupation of the low-energy modes is evident.

%\js{\cs\cs EXPERIMENT 2: HI-RES SPECTROSCOPY, BEC}

\begin{figure}[t]
    \centering
    \includegraphics[width=1.0\columnwidth]{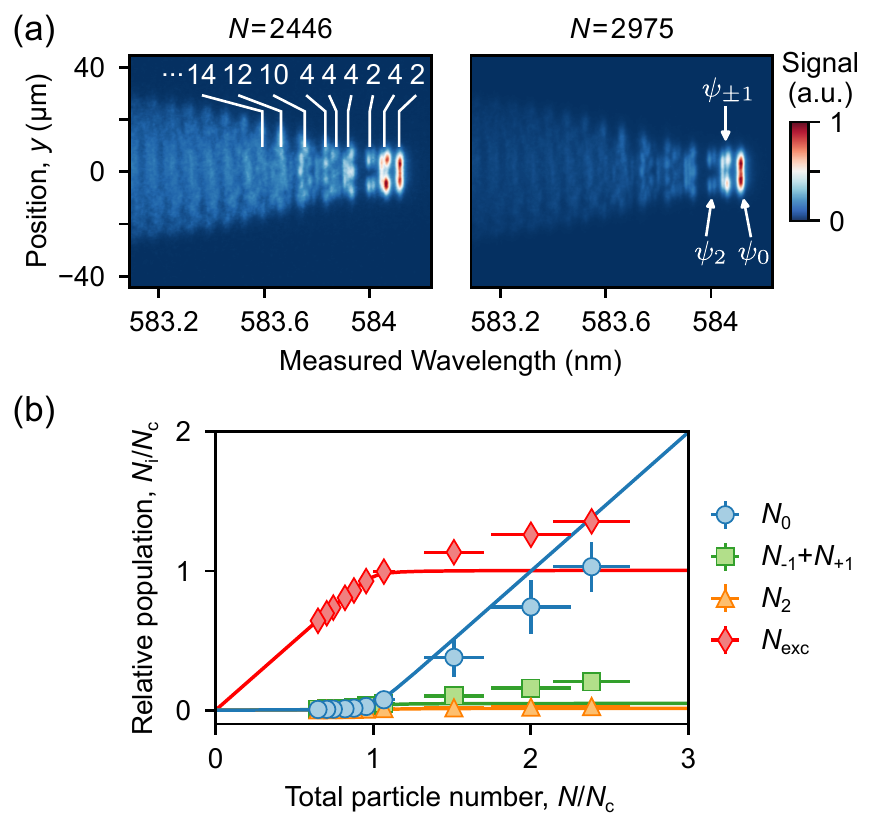}
\caption{(a)~Spectrally highly resolved data (normalized) showing the distribution of the microcavity emission as a function of the measured wavelength and position along the $y$-axis, corresponding to one of the lattice axes, for $N/\Nc = \SI{0.88 \pm 0.03}{}$ (left panel) and $N/\Nc = \SI{1.07 \pm 0.07}{}$ (right); numbers in the left panel give the mode degeneracy determined independently by exciting the transverse modes with a focused laser beam. The three rightmost spectral lines (at largest wavelengths) correspond to the four ring modes (see arrows), while signals at smaller wavelengths correspond to harmonic oscillator ones. For the large photon number, the Bose-Einstein condensate is seen in the increased emission from the lowest energy state. (b)~Relative population in the resolved ring eigenstates $N_0$, $N_{-1}+N_{+1}$, and $N_{2}$, as well as in all available excited states $N_\mathrm{exc}=N-N_0$ (symbols) versus total photon number along with theory (lines). All numbers are normalized by $\Nc$. Error bars give standard statistical errors; the cutoff wavelength is at $\SI{584.1}{\nano\meter}$.}
    \label{fig:2}
\end{figure}

In order to shed more light onto the observed macroscopic occupation, we recorded spectral distributions of the low-energy photons using a highly-resolving spectrometer (resolution $\SI{20}{\giga\hertz}$); the used spectrometer allows us to resolve the lower energetic part of the transverse mode spectrum and extract the state populations. Figure~\ref{fig:2}(a) shows typical experimental data for photon numbers below and above the Bose-Einstein condensation threshold, respectively. As we use a slitless spectrometer, the dispersed cavity emission still contains information on the spatial mode profile. Along the spectrally dispersed direction ($x$-axis), a cylindrical telescope with a 6:1 aspect ratio reduces the imaging scale correspondingly. The spatial mode profile along the $y$-axis is visible, for example, for the longest wavelength (lowest photon energy) mode with a profile resembling the symmetric ground state $\psi_0$ of the ring system, while in the spatial distribution of the third mode a pronounced minimum near $y = 0$ (as well as a minimum along $x$) is observed, as expected for the antisymmetric state $\psi_2$. The signal observed at the intermediate wavelength is attributed to the degenerate states $\psi_{\pm 1}$, and the observed spectrum confirms their degeneracy within our experimental accuracy. The measured frequency splitting of the hybridized eigenstates $\psi_0$ and $\psi_{\pm 1}$ or $\psi_{\pm 1}$ and $\psi_2$ is \SI{48.5 \pm 0.4}{\giga\hertz} and \SI{50.4 \pm 0.9}{\giga\hertz}, respectively. These values are below the theoretically expected splittings for the mirror surface profile, which we have determined independently using white-light interferometry~\cite{Kurtscheid:2020, Supplementary}. For the next higher energetic modes, we find that their wavelengths and degeneracies cannot be fully described by harmonic oscillator modes. This behavior is well understood from the influence of the lattice potential still being relevant for the energetically lowest-lying harmonic oscillator states. For energies of more than $h \cdot \SI{400}{\giga\hertz}$ above the cutoff, the levels become essentially equidistantly spaced and the mode degeneracy increases linearly, corresponding to the expectations for the harmonic oscillator case of $g(u) = 2(1 + u/\hbar \Omega)$, where $u$ denotes the energy above the cutoff.

%%%%%%%%%%%%%%%%%
%%%%% FIG 2 %%%%%
%%%%%%%%%%%%%%%%%

%\js{\cs EXPERIMENT 3: RING OCCUPATION + THERMODYNAMICS}

To obtain quantitative insight into the behavior at the phase transition, Fig.~\ref{fig:2}(b) shows the relative population in the low-energy eigenstates of the ring lattice for different photon numbers. Around $\Nc$ the ground state population $N_0$ features a clear threshold behavior, which is revealed by a sudden increase of $N_0$ with increasing total particle number. Due to the finite size of the system, this phase transition is softened, and correspondingly we determined $\Nc$ by a linear fit to the data at large photon numbers. The saturation of the excited mode population $N_\mathrm{exc}$ expected for a Bose-Einstein condensate is to good approximation fulfilled. Indeed, the dominant contribution to the residual growth of $N_\mathrm{exc}$ in the condensed regime results largely from the degenerate pair of lattice modes $\psi_{\pm1}$, as is clearly observed in Fig.~\ref{fig:2}(b). This behavior is attributed mainly to an incomplete equilibration of the photon gas due to the driven-dissipative nature of the system~\cite{Keeling:2016}.

%%%%%%%%%%%%%%%%%
%%%%% FIG 3 %%%%%
%%%%%%%%%%%%%%%%%

%\js{\cs\cs EXPERIMENT 4: SPATIAL DISTRIBUTION}

The condensation is also revealed in the spatial distribution of the photon gas obtained by imaging the emission from the microcavity plane onto a camera. Experimental data for photon numbers corresponding to below and above $\Nc$ is shown in Fig.~\ref{fig:3}(a). For the data taken in the uncondensed regime [Fig.~\ref{fig:3}(a), left panel], we observe a broad thermal distribution up to the high-energy edge of the imprinted harmonic potential, which is radially located in $\SI{60}{\micro\meter}$ distance from the trap center. Above the critical photon number [Fig.~\ref{fig:3}(a), right panel], we observe a four-peaked intensity distribution in the center on top of the thermal cloud, indicating the macroscopic occupation of photons in the symmetric superposition state $\psi_0$ upon Bose-Einstein condensation. The spatial intensity profile of the populated ground state shows a slight asymmetry among the four lattice sites, which is attributed to residual detunings between the bare eigenenergies caused by height variations of the imprinted potentials below $\SI{1}{\angstrom}$~\cite{Kurtscheid:2020}. Figure~\ref{fig:3}(b) shows corresponding line densities obtained from integration along the vertical axes in Fig.~\ref{fig:3}(a), along with theory based on a Bose-Einstein distributed occupation of the modes bound in the finite-depth potential. The deviation between the measured photon numbers and the fitted ones are attributed to an imperfect saturation of the excited mode population [as previously seen in Fig.~\ref{fig:2}(b)]. Following the spectroscopic confirmation, the agreement of the measured spatial profiles with the theory prediction gives a second line of evidence for the thermalization of the two-dimensional photon gas.

%%%%%%%%%%%%%%%%%
%%%%% FIG 4 %%%%%
%%%%%%%%%%%%%%%%%
\begin{figure}[t]
    \centering
    \includegraphics[width=1.0\columnwidth]{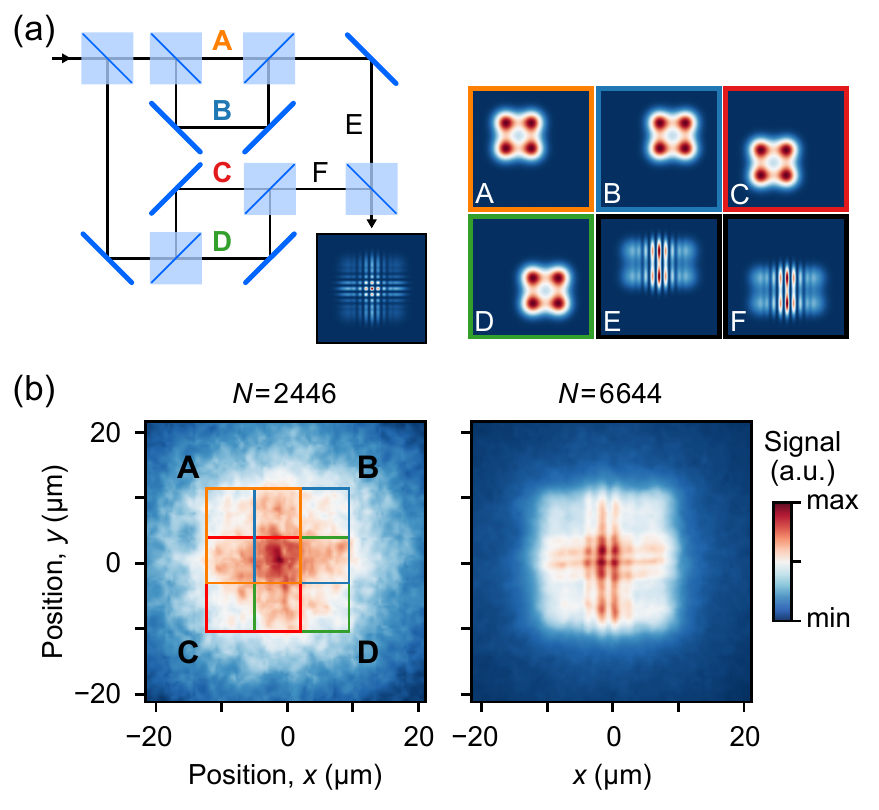}
\caption{(a)~Interferometry setup to probe the mutual phase coherence of photons from different lattice sites. By spatially overlapping the fields emitted from different transverse points in the cavity, interference signals between two and four sites are recorded. The images on the right show calculated intensity contributions (for pure condensates) from the individual paths A-D, as well as their expected intensity along path E and F. (b)~Interference signal below (left) and above (right) the critical photon number for Bose-Einstein condensation with $N/\Nc = \SI{0.88 \pm 0.03}{}$ and $N/\Nc = \SI{2.4 \pm 0.2}{}$, respectively. As in (a), the boxes indicate how the four paths A-D are overlapped in the detector plane; note that the corresponding wave vectors are deliberately tilted to obtain a visible fringe pattern. For the data recorded in the condensed phase, a stable interference signal is visible in regions where the emission from different sites overlap.}
    \label{fig:4}
\end{figure}

%\js{\cs\cs EXPERIMENT 5: INTERFEROMETRY, PHASE COHERENCE}

The four hybridized eigenstates can be uniquely characterized by their corresponding relative phase between the localized lattice site eigenfunctions. To verify the expected phase coherence among the different lattice sites, we perform optical interferometry of the cavity emission. As shown in Fig.~\ref{fig:4}(a), the cavity output was split into four paths and subsequently recombined, each offset by approximately half the image size, giving access to two-, and four-site interference in different spatial regions of the recorded interference pattern. Figure~\ref{fig:4}(b) shows typical false-color images obtained by imaging the interferometer output on a camera for $N<\Nc$ (left panel) and $N>\Nc$ (right), respectively. While below the condensation threshold no interference signal is visible, above the condensation threshold a clear interference signal is visible in the regions where the emission of two or more sites spatially overlap. Given that the image is the average of data recorded in many subsequent realizations of the experiment, this confirms the existence of a stable phase relation between the optical fields at different sites. The presence of this fringe pattern confirms their coherence, indicating that a macroscopic wave function extends over all four lattice sites; in other words, a photon is delocalized over the entire ring lattice structure. The visible interference contrast here sets an upper limit to the “which-path” information (here better termed as “which-site information”) occuring in the process of thermalization by contact to the dye molecules. For the parameters used in our experiment, the characteristic timescale for photons moving between sites ($h \pi / (2J) \approx \SI{10}{\pico\second}$) is much shorter than that for dye-mediated thermalization $\approx \SI{130}{\pico\second}$~\cite{Schmitt:2015}, so that the presence of a fringe pattern is well within expectations even in the presence of the coupling to a thermal bath.

%\js{\cs\cs CONCLUSION}

To conclude, we have demonstrated Bose-Einstein condensation of photons into the hybridized ground state of a ring lattice. The observed spatial and spectral distributions are in good agreement with equilibrium theory, confirming that the macroscopic state preparation is driven by quantum statistics, in contrast to classical optical state preparation methods where photon loss is balanced by gain. Further, the fixed phase relation of the photons at the different sites of the ring lattice was verified interferometrically. 

For the future, lattices of interacting photons induced, \emph{e.g.}, by effective Kerr nonlinearities offer prospects for the preparation of entangled many-body ground states using the demonstrated thermal equilibrium process~\cite{Kurtscheid:2019,Majumdar:2013}. Combining this approach with reversible thermo-optic methods will enable also the control of spatially extended lattice systems~\cite{Dung:2017, Vretenar:2021b, Busley:2022}. The realized quantum ring furthermore holds promise for the engineering of optical analogs of superconducting flux qubits~\cite{Xue:2021}. Other prospects include the "cooling" into low-energy topological states of light based on circular modulations of the tunneling~\cite{Fang:2012}, or in a classical regime, simulations of the XY-model~\cite{Berloff:2017,Gladilin:2020b}.

%(leading to which-way information) 

\begin{acknowledgments}
We acknowledge financial support by the DFG within SFB/TR185 (277625399) and the Cluster of Excellence ML4Q (EXC 2004/1-390534769), and the DLR with funds provided by the BMWK (50WM2240). J.S. acknowledges support by the EU (ERC, TopoGrand, 101040409).
\end{acknowledgments}


\begin{thebibliography}{47}%
	\makeatletter
	\providecommand \@ifxundefined [1]{%
		\@ifx{#1\undefined}
	}%
	\providecommand \@ifnum [1]{%
		\ifnum #1\expandafter \@firstoftwo
		\else \expandafter \@secondoftwo
		\fi
	}%
	\providecommand \@ifx [1]{%
		\ifx #1\expandafter \@firstoftwo
		\else \expandafter \@secondoftwo
		\fi
	}%
	\providecommand \natexlab [1]{#1}%
	\providecommand \enquote  [1]{``#1''}%
	\providecommand \bibnamefont  [1]{#1}%
	\providecommand \bibfnamefont [1]{#1}%
	\providecommand \citenamefont [1]{#1}%
	\providecommand \href@noop [0]{\@secondoftwo}%
	\providecommand \href [0]{\begingroup \@sanitize@url \@href}%
	\providecommand \@href[1]{\@@startlink{#1}\@@href}%
	\providecommand \@@href[1]{\endgroup#1\@@endlink}%
	\providecommand \@sanitize@url [0]{\catcode `\\12\catcode `\$12\catcode
		`\&12\catcode `\#12\catcode `\^12\catcode `\_12\catcode `\%12\relax}%
	\providecommand \@@startlink[1]{}%
	\providecommand \@@endlink[0]{}%
	\providecommand \url  [0]{\begingroup\@sanitize@url \@url }%
	\providecommand \@url [1]{\endgroup\@href {#1}{\urlprefix }}%
	\providecommand \urlprefix  [0]{URL }%
	\providecommand \Eprint [0]{\href }%
	\providecommand \doibase [0]{https://doi.org/}%
	\providecommand \selectlanguage [0]{\@gobble}%
	\providecommand \bibinfo  [0]{\@secondoftwo}%
	\providecommand \bibfield  [0]{\@secondoftwo}%
	\providecommand \translation [1]{[#1]}%
	\providecommand \BibitemOpen [0]{}%
	\providecommand \bibitemStop [0]{}%
	\providecommand \bibitemNoStop [0]{.\EOS\space}%
	\providecommand \EOS [0]{\spacefactor3000\relax}%
	\providecommand \BibitemShut  [1]{\csname bibitem#1\endcsname}%
	\let\auto@bib@innerbib\@empty
	%</preamble>
	\bibitem [{\citenamefont {Chang}(2003)}]{Chang:2003}%
	\BibitemOpen
	\bibfield  {author} {\bibinfo {author} {\bibfnamefont {A.~M.}\ \bibnamefont
			{Chang}},\ }\bibfield  {title} {\bibinfo {title} {Chiral {L}uttinger liquids
			at the fractional quantum {H}all edge},\ }\href
	{https://doi.org/10.1103/RevModPhys.75.1449} {\bibfield  {journal} {\bibinfo
			{journal} {Rev. Mod. Phys.}\ }\textbf {\bibinfo {volume} {75}},\ \bibinfo
		{pages} {1449} (\bibinfo {year} {2003})}\BibitemShut {NoStop}%
	\bibitem [{\citenamefont {Deshpande}\ \emph {et~al.}(2010)\citenamefont
		{Deshpande}, \citenamefont {Bockrath}, \citenamefont {Glazman},\ and\
		\citenamefont {Yacoby}}]{Deshpande:2010}%
	\BibitemOpen
	\bibfield  {author} {\bibinfo {author} {\bibfnamefont {V.~V.}\ \bibnamefont
			{Deshpande}}, \bibinfo {author} {\bibfnamefont {M.}~\bibnamefont {Bockrath}},
		\bibinfo {author} {\bibfnamefont {L.~I.}\ \bibnamefont {Glazman}},\ and\
		\bibinfo {author} {\bibfnamefont {A.}~\bibnamefont {Yacoby}},\ }\bibfield
	{title} {\bibinfo {title} {Electron liquids and solids in one dimension},\
	}\href {https://doi.org/10.1038/nature08918} {\bibfield  {journal} {\bibinfo
			{journal} {Nature}\ }\textbf {\bibinfo {volume} {464}},\ \bibinfo {pages}
		{209} (\bibinfo {year} {2010})}\BibitemShut {NoStop}%
	\bibitem [{\citenamefont {Greiner}\ \emph {et~al.}(2002)\citenamefont
		{Greiner}, \citenamefont {Mandel}, \citenamefont {Esslinger}, \citenamefont
		{H{\"a}nsch},\ and\ \citenamefont {Bloch}}]{Greiner:2002a}%
	\BibitemOpen
	\bibfield  {author} {\bibinfo {author} {\bibfnamefont {M.}~\bibnamefont
			{Greiner}}, \bibinfo {author} {\bibfnamefont {O.}~\bibnamefont {Mandel}},
		\bibinfo {author} {\bibfnamefont {T.}~\bibnamefont {Esslinger}}, \bibinfo
		{author} {\bibfnamefont {T.~W.}\ \bibnamefont {H{\"a}nsch}},\ and\ \bibinfo
		{author} {\bibfnamefont {I.}~\bibnamefont {Bloch}},\ }\bibfield  {title}
	{\bibinfo {title} {Quantum phase transition from a superfluid to a {M}ott
			insulator in a gas of ultracold atoms},\ }\href
	{https://doi.org/10.1038/415039a} {\bibfield  {journal} {\bibinfo  {journal}
			{Nature}\ }\textbf {\bibinfo {volume} {415}},\ \bibinfo {pages} {39}
		(\bibinfo {year} {2002})}\BibitemShut {NoStop}%
	\bibitem [{\citenamefont {Bao}\ \emph {et~al.}(2004)\citenamefont {Bao},
		\citenamefont {Kim}, \citenamefont {Mawst}, \citenamefont {Elkin},
		\citenamefont {Troshchieva}, \citenamefont {Vysotsky},\ and\ \citenamefont
		{Napartovich}}]{Bao:2004}%
	\BibitemOpen
	\bibfield  {author} {\bibinfo {author} {\bibfnamefont {L.}~\bibnamefont
			{Bao}}, \bibinfo {author} {\bibfnamefont {N.-H.}\ \bibnamefont {Kim}},
		\bibinfo {author} {\bibfnamefont {L.~J.}\ \bibnamefont {Mawst}}, \bibinfo
		{author} {\bibfnamefont {N.~N.}\ \bibnamefont {Elkin}}, \bibinfo {author}
		{\bibfnamefont {V.~N.}\ \bibnamefont {Troshchieva}}, \bibinfo {author}
		{\bibfnamefont {D.~V.}\ \bibnamefont {Vysotsky}},\ and\ \bibinfo {author}
		{\bibfnamefont {A.~P.}\ \bibnamefont {Napartovich}},\ }\bibfield  {title}
	{\bibinfo {title} {Near-diffraction-limited coherent emission from large
			aperture antiguided vertical-cavity surface-emitting laser arrays},\ }\href
	{https://doi.org/10.1063/1.1640799} {\bibfield  {journal} {\bibinfo
			{journal} {Appl. Phys. Lett.}\ }\textbf {\bibinfo {volume} {84}},\ \bibinfo
		{pages} {320} (\bibinfo {year} {2004})}\BibitemShut {NoStop}%
	\bibitem [{\citenamefont {Nixon}\ \emph {et~al.}(2013)\citenamefont {Nixon},
		\citenamefont {Ronen}, \citenamefont {Friesem},\ and\ \citenamefont
		{Davidson}}]{Nixon:2013}%
	\BibitemOpen
	\bibfield  {author} {\bibinfo {author} {\bibfnamefont {M.}~\bibnamefont
			{Nixon}}, \bibinfo {author} {\bibfnamefont {E.}~\bibnamefont {Ronen}},
		\bibinfo {author} {\bibfnamefont {A.~A.}\ \bibnamefont {Friesem}},\ and\
		\bibinfo {author} {\bibfnamefont {N.}~\bibnamefont {Davidson}},\ }\bibfield
	{title} {\bibinfo {title} {Observing geometric frustration with thousands of
			coupled lasers},\ }\href {https://doi.org/10.1103/PhysRevLett.110.184102}
	{\bibfield  {journal} {\bibinfo  {journal} {Phys. Rev. Lett.}\ }\textbf
		{\bibinfo {volume} {110}},\ \bibinfo {pages} {184102} (\bibinfo {year}
		{2013})}\BibitemShut {NoStop}%
	\bibitem [{\citenamefont {Deng}\ \emph {et~al.}(2010)\citenamefont {Deng},
		\citenamefont {Haug},\ and\ \citenamefont {Yamamoto}}]{Deng:2010}%
	\BibitemOpen
	\bibfield  {author} {\bibinfo {author} {\bibfnamefont {H.}~\bibnamefont
			{Deng}}, \bibinfo {author} {\bibfnamefont {H.}~\bibnamefont {Haug}},\ and\
		\bibinfo {author} {\bibfnamefont {Y.}~\bibnamefont {Yamamoto}},\ }\bibfield
	{title} {\bibinfo {title} {Exciton-polariton {B}ose-{E}instein
			condensation},\ }\href {https://doi.org/10.1103/RevModPhys.82.1489}
	{\bibfield  {journal} {\bibinfo  {journal} {Rev. Mod. Phys.}\ }\textbf
		{\bibinfo {volume} {82}},\ \bibinfo {pages} {1489} (\bibinfo {year}
		{2010})}\BibitemShut {NoStop}%
	\bibitem [{\citenamefont {Carusotto}\ and\ \citenamefont
		{Ciuti}(2013)}]{Carusotto:2013}%
	\BibitemOpen
	\bibfield  {author} {\bibinfo {author} {\bibfnamefont {I.}~\bibnamefont
			{Carusotto}}\ and\ \bibinfo {author} {\bibfnamefont {C.}~\bibnamefont
			{Ciuti}},\ }\bibfield  {title} {\bibinfo {title} {Quantum fluids of light},\
	}\href {https://doi.org/10.1103/RevModPhys.85.299} {\bibfield  {journal}
		{\bibinfo  {journal} {Rev. Mod. Phys.}\ }\textbf {\bibinfo {volume} {85}},\
		\bibinfo {pages} {299} (\bibinfo {year} {2013})}\BibitemShut {NoStop}%
	\bibitem [{\citenamefont {Klaers}\ \emph {et~al.}(2010)\citenamefont {Klaers},
		\citenamefont {Schmitt}, \citenamefont {Vewinger},\ and\ \citenamefont
		{Weitz}}]{Klaers:2010}%
	\BibitemOpen
	\bibfield  {author} {\bibinfo {author} {\bibfnamefont {J.}~\bibnamefont
			{Klaers}}, \bibinfo {author} {\bibfnamefont {J.}~\bibnamefont {Schmitt}},
		\bibinfo {author} {\bibfnamefont {F.}~\bibnamefont {Vewinger}},\ and\
		\bibinfo {author} {\bibfnamefont {M.}~\bibnamefont {Weitz}},\ }\bibfield
	{title} {\bibinfo {title} {{B}ose--{E}instein condensation of photons in an
			optical microcavity},\ }\href {https://doi.org/10.1038/nature09567}
	{\bibfield  {journal} {\bibinfo  {journal} {Nature}\ }\textbf {\bibinfo
			{volume} {468}},\ \bibinfo {pages} {545} (\bibinfo {year}
		{2010})}\BibitemShut {NoStop}%
	\bibitem [{\citenamefont {Schmitt}(2018)}]{Schmitt:2018}%
	\BibitemOpen
	\bibfield  {author} {\bibinfo {author} {\bibfnamefont {J.}~\bibnamefont
			{Schmitt}},\ }\bibfield  {title} {\bibinfo {title} {Dynamics and correlations
			of a {B}ose{\textendash}{E}instein condensate of photons},\ }\href
	{https://doi.org/10.1088/1361-6455/aad409} {\bibfield  {journal} {\bibinfo
			{journal} {J. Phys. B: At. Mol. Opt. Phys.}\ }\textbf {\bibinfo {volume}
			{51}},\ \bibinfo {pages} {173001} (\bibinfo {year} {2018})}\BibitemShut
	{NoStop}%
	\bibitem [{\citenamefont {Bloch}\ \emph {et~al.}(2022)\citenamefont {Bloch},
		\citenamefont {Carusotto},\ and\ \citenamefont {Wouters}}]{Bloch:2022}%
	\BibitemOpen
	\bibfield  {author} {\bibinfo {author} {\bibfnamefont {J.}~\bibnamefont
			{Bloch}}, \bibinfo {author} {\bibfnamefont {I.}~\bibnamefont {Carusotto}},\
		and\ \bibinfo {author} {\bibfnamefont {M.}~\bibnamefont {Wouters}},\
	}\bibfield  {title} {\bibinfo {title} {Non-equilibrium {B}ose--{E}instein
			condensation in photonic systems},\ }\href
	{https://doi.org/10.1038/s42254-022-00464-0} {\bibfield  {journal} {\bibinfo
			{journal} {Nat. Rev. Phys.}\ }\textbf {\bibinfo {volume} {4}},\ \bibinfo
		{pages} {470–488} (\bibinfo {year} {2022})}\BibitemShut {NoStop}%
	\bibitem [{\citenamefont {Marelic}\ and\ \citenamefont
		{Nyman}(2015)}]{Marelic:2015}%
	\BibitemOpen
	\bibfield  {author} {\bibinfo {author} {\bibfnamefont {J.}~\bibnamefont
			{Marelic}}\ and\ \bibinfo {author} {\bibfnamefont {R.~A.}\ \bibnamefont
			{Nyman}},\ }\bibfield  {title} {\bibinfo {title} {Experimental evidence for
			inhomogeneous pumping and energy-dependent effects in photon
			{B}ose-{E}instein condensation},\ }\href
	{https://doi.org/10.1103/PhysRevA.91.033813} {\bibfield  {journal} {\bibinfo
			{journal} {Phys. Rev. A}\ }\textbf {\bibinfo {volume} {91}},\ \bibinfo
		{pages} {033813} (\bibinfo {year} {2015})}\BibitemShut {NoStop}%
	\bibitem [{\citenamefont {Greveling}\ \emph {et~al.}(2018)\citenamefont
		{Greveling}, \citenamefont {Perrier},\ and\ \citenamefont {van
			Oosten}}]{Greveling:2018}%
	\BibitemOpen
	\bibfield  {author} {\bibinfo {author} {\bibfnamefont {S.}~\bibnamefont
			{Greveling}}, \bibinfo {author} {\bibfnamefont {K.~L.}\ \bibnamefont
			{Perrier}},\ and\ \bibinfo {author} {\bibfnamefont {D.}~\bibnamefont {van
				Oosten}},\ }\bibfield  {title} {\bibinfo {title} {Density distribution of a
			{B}ose-{E}instein condensate of photons in a dye-filled microcavity},\ }\href
	{https://doi.org/10.1103/PhysRevA.98.013810} {\bibfield  {journal} {\bibinfo
			{journal} {Phys. Rev. A}\ }\textbf {\bibinfo {volume} {98}},\ \bibinfo
		{pages} {013810} (\bibinfo {year} {2018})}\BibitemShut {NoStop}%
	\bibitem [{\citenamefont {Vretenar}\ \emph
		{et~al.}(2021{\natexlab{a}})\citenamefont {Vretenar}, \citenamefont
		{Toebes},\ and\ \citenamefont {Klaers}}]{Vretenar:2021a}%
	\BibitemOpen
	\bibfield  {author} {\bibinfo {author} {\bibfnamefont {M.}~\bibnamefont
			{Vretenar}}, \bibinfo {author} {\bibfnamefont {C.}~\bibnamefont {Toebes}},\
		and\ \bibinfo {author} {\bibfnamefont {J.}~\bibnamefont {Klaers}},\
	}\bibfield  {title} {\bibinfo {title} {Modified {B}ose-{E}instein
			condensation in an optical quantum gas},\ }\href
	{https://doi.org/10.1038/s41467-021-26087-0} {\bibfield  {journal} {\bibinfo
			{journal} {Nat. Comm.}\ }\textbf {\bibinfo {volume} {12}},\ \bibinfo {pages}
		{5749} (\bibinfo {year} {2021}{\natexlab{a}})}\BibitemShut {NoStop}%
	\bibitem [{\citenamefont {Weill}\ \emph {et~al.}(2019)\citenamefont {Weill},
		\citenamefont {Bekker}, \citenamefont {Levit},\ and\ \citenamefont
		{Fischer}}]{Weill:2019}%
	\BibitemOpen
	\bibfield  {author} {\bibinfo {author} {\bibfnamefont {R.}~\bibnamefont
			{Weill}}, \bibinfo {author} {\bibfnamefont {A.}~\bibnamefont {Bekker}},
		\bibinfo {author} {\bibfnamefont {B.}~\bibnamefont {Levit}},\ and\ \bibinfo
		{author} {\bibfnamefont {B.}~\bibnamefont {Fischer}},\ }\bibfield  {title}
	{\bibinfo {title} {Bose--{E}instein condensation of photons in an
			erbium--ytterbium co-doped fiber cavity},\ }\href
	{https://doi.org/10.1038/s41467-019-08527-0} {\bibfield  {journal} {\bibinfo
			{journal} {Nat. Comm.}\ }\textbf {\bibinfo {volume} {10}},\ \bibinfo {pages}
		{747} (\bibinfo {year} {2019})}\BibitemShut {NoStop}%
	\bibitem [{\citenamefont {Hakala}\ \emph {et~al.}(2018)\citenamefont {Hakala},
		\citenamefont {Moilanen}, \citenamefont {V{\"a}kev{\"a}inen}, \citenamefont
		{Guo}, \citenamefont {Martikainen}, \citenamefont {Daskalakis}, \citenamefont
		{Rekola}, \citenamefont {Julku},\ and\ \citenamefont
		{T{\"o}rm{\"a}}}]{Hakala:2018}%
	\BibitemOpen
	\bibfield  {author} {\bibinfo {author} {\bibfnamefont {T.~K.}\ \bibnamefont
			{Hakala}}, \bibinfo {author} {\bibfnamefont {A.~J.}\ \bibnamefont
			{Moilanen}}, \bibinfo {author} {\bibfnamefont {A.~I.}\ \bibnamefont
			{V{\"a}kev{\"a}inen}}, \bibinfo {author} {\bibfnamefont {R.}~\bibnamefont
			{Guo}}, \bibinfo {author} {\bibfnamefont {J.-P.}\ \bibnamefont
			{Martikainen}}, \bibinfo {author} {\bibfnamefont {K.~S.}\ \bibnamefont
			{Daskalakis}}, \bibinfo {author} {\bibfnamefont {H.~T.}\ \bibnamefont
			{Rekola}}, \bibinfo {author} {\bibfnamefont {A.}~\bibnamefont {Julku}},\ and\
		\bibinfo {author} {\bibfnamefont {P.}~\bibnamefont {T{\"o}rm{\"a}}},\
	}\bibfield  {title} {\bibinfo {title} {{B}ose--{E}instein condensation in a
			plasmonic lattice},\ }\href {https://doi.org/10.1038/s41567-018-0109-9}
	{\bibfield  {journal} {\bibinfo  {journal} {Nat. Phys.}\ }\textbf {\bibinfo
			{volume} {14}},\ \bibinfo {pages} {739} (\bibinfo {year} {2018})}\BibitemShut
	{NoStop}%
	\bibitem [{\citenamefont {Schofield}\ \emph {et~al.}(2024)\citenamefont
		{Schofield}, \citenamefont {Fu}, \citenamefont {Clarke}, \citenamefont
		{Farrer}, \citenamefont {Trapalis}, \citenamefont {Dhar}, \citenamefont
		{Mukherjee}, \citenamefont {Severs~Millard}, \citenamefont {Heffernan},
		\citenamefont {Mintert} \emph {et~al.}}]{Schofield:2024}%
	\BibitemOpen
	\bibfield  {author} {\bibinfo {author} {\bibfnamefont {R.~C.}\ \bibnamefont
			{Schofield}}, \bibinfo {author} {\bibfnamefont {M.}~\bibnamefont {Fu}},
		\bibinfo {author} {\bibfnamefont {E.}~\bibnamefont {Clarke}}, \bibinfo
		{author} {\bibfnamefont {I.}~\bibnamefont {Farrer}}, \bibinfo {author}
		{\bibfnamefont {A.}~\bibnamefont {Trapalis}}, \bibinfo {author}
		{\bibfnamefont {H.~S.}\ \bibnamefont {Dhar}}, \bibinfo {author}
		{\bibfnamefont {R.}~\bibnamefont {Mukherjee}}, \bibinfo {author}
		{\bibfnamefont {T.}~\bibnamefont {Severs~Millard}}, \bibinfo {author}
		{\bibfnamefont {J.}~\bibnamefont {Heffernan}}, \bibinfo {author}
		{\bibfnamefont {F.}~\bibnamefont {Mintert}}, \emph {et~al.},\ }\bibfield
	{title} {\bibinfo {title} {Bose--einstein condensation of light in a
			semiconductor quantum well microcavity},\ }\href
	{https://doi.org/10.1038/s41566-024-01491-2} {\bibfield  {journal} {\bibinfo
			{journal} {Nature Photonics}\ ,\ \bibinfo {pages} {1}} (\bibinfo {year}
		{2024})}\BibitemShut {NoStop}%
	\bibitem [{\citenamefont {Pieczarka}\ \emph {et~al.}(2024)\citenamefont
		{Pieczarka}, \citenamefont {G{\k{e}}bski}, \citenamefont {Piasecka},
		\citenamefont {Lott}, \citenamefont {Pelster}, \citenamefont {Wasiak},\ and\
		\citenamefont {Czyszanowski}}]{Pieczarka:2024}%
	\BibitemOpen
	\bibfield  {author} {\bibinfo {author} {\bibfnamefont {M.}~\bibnamefont
			{Pieczarka}}, \bibinfo {author} {\bibfnamefont {M.}~\bibnamefont
			{G{\k{e}}bski}}, \bibinfo {author} {\bibfnamefont {A.~N.}\ \bibnamefont
			{Piasecka}}, \bibinfo {author} {\bibfnamefont {J.~A.}\ \bibnamefont {Lott}},
		\bibinfo {author} {\bibfnamefont {A.}~\bibnamefont {Pelster}}, \bibinfo
		{author} {\bibfnamefont {M.}~\bibnamefont {Wasiak}},\ and\ \bibinfo {author}
		{\bibfnamefont {T.}~\bibnamefont {Czyszanowski}},\ }\bibfield  {title}
	{\bibinfo {title} {Bose--einstein condensation of photons in a
			vertical-cavity surface-emitting laser},\ }\href
	{https://doi.org/10.1038/s41566-024-01478-z} {\bibfield  {journal} {\bibinfo
			{journal} {Nature Photonics}\ ,\ \bibinfo {pages} {1}} (\bibinfo {year}
		{2024})}\BibitemShut {NoStop}%
	\bibitem [{\citenamefont {Figueiredo}\ \emph {et~al.}(2023)\citenamefont
		{Figueiredo}, \citenamefont {Mendon\ifmmode~\mbox{\c{c}}\else \c{c}\fi{}a},\
		and\ \citenamefont {Ter\ifmmode~\mbox{\c{c}}\else
			\c{c}\fi{}as}}]{Figueiredo:2023}%
	\BibitemOpen
	\bibfield  {author} {\bibinfo {author} {\bibfnamefont {J.~L.}\ \bibnamefont
			{Figueiredo}}, \bibinfo {author} {\bibfnamefont {J.~T.}\ \bibnamefont
			{Mendon\ifmmode~\mbox{\c{c}}\else \c{c}\fi{}a}},\ and\ \bibinfo {author}
		{\bibfnamefont {H.}~\bibnamefont {Ter\ifmmode~\mbox{\c{c}}\else
				\c{c}\fi{}as}},\ }\bibfield  {title} {\bibinfo {title} {Bose-einstein
			condensation of photons in microcavity plasmas},\ }\href
	{https://doi.org/10.1103/PhysRevE.108.L013201} {\bibfield  {journal}
		{\bibinfo  {journal} {Phys. Rev. E}\ }\textbf {\bibinfo {volume} {108}},\
		\bibinfo {pages} {L013201} (\bibinfo {year} {2023})}\BibitemShut {NoStop}%
	\bibitem [{\citenamefont {Klaers}\ \emph {et~al.}(2012)\citenamefont {Klaers},
		\citenamefont {Schmitt}, \citenamefont {Damm}, \citenamefont {Vewinger},\
		and\ \citenamefont {Weitz}}]{Klaers:2012}%
	\BibitemOpen
	\bibfield  {author} {\bibinfo {author} {\bibfnamefont {J.}~\bibnamefont
			{Klaers}}, \bibinfo {author} {\bibfnamefont {J.}~\bibnamefont {Schmitt}},
		\bibinfo {author} {\bibfnamefont {T.}~\bibnamefont {Damm}}, \bibinfo {author}
		{\bibfnamefont {F.}~\bibnamefont {Vewinger}},\ and\ \bibinfo {author}
		{\bibfnamefont {M.}~\bibnamefont {Weitz}},\ }\bibfield  {title} {\bibinfo
		{title} {Statistical physics of {B}ose-{E}instein-condensed light in a dye
			microcavity},\ }\href {https://doi.org/10.1103/PhysRevLett.108.160403}
	{\bibfield  {journal} {\bibinfo  {journal} {Phys. Rev. Lett.}\ }\textbf
		{\bibinfo {volume} {108}},\ \bibinfo {pages} {160403} (\bibinfo {year}
		{2012})}\BibitemShut {NoStop}%
	\bibitem [{\citenamefont {Berloff}\ \emph {et~al.}(2017)\citenamefont
		{Berloff}, \citenamefont {Silva}, \citenamefont {Kalinin}, \citenamefont
		{Askitopoulos}, \citenamefont {T{\"o}pfer}, \citenamefont {Cilibrizzi},
		\citenamefont {Langbein},\ and\ \citenamefont {Lagoudakis}}]{Berloff:2017}%
	\BibitemOpen
	\bibfield  {author} {\bibinfo {author} {\bibfnamefont {N.~G.}\ \bibnamefont
			{Berloff}}, \bibinfo {author} {\bibfnamefont {M.}~\bibnamefont {Silva}},
		\bibinfo {author} {\bibfnamefont {K.}~\bibnamefont {Kalinin}}, \bibinfo
		{author} {\bibfnamefont {A.}~\bibnamefont {Askitopoulos}}, \bibinfo {author}
		{\bibfnamefont {J.~D.}\ \bibnamefont {T{\"o}pfer}}, \bibinfo {author}
		{\bibfnamefont {P.}~\bibnamefont {Cilibrizzi}}, \bibinfo {author}
		{\bibfnamefont {W.}~\bibnamefont {Langbein}},\ and\ \bibinfo {author}
		{\bibfnamefont {P.~G.}\ \bibnamefont {Lagoudakis}},\ }\bibfield  {title}
	{\bibinfo {title} {Realizing the classical {XY} {H}amiltonian in polariton
			simulators},\ }\href {https://doi.org/10.1038/nmat4971} {\bibfield  {journal}
		{\bibinfo  {journal} {Nat. Mater.}\ }\textbf {\bibinfo {volume} {16}},\
		\bibinfo {pages} {1120} (\bibinfo {year} {2017})}\BibitemShut {NoStop}%
	\bibitem [{\citenamefont {Gershenzon}\ \emph {et~al.}(2020)\citenamefont
		{Gershenzon}, \citenamefont {Arwas}, \citenamefont {Gadasi}, \citenamefont
		{Tradonsky}, \citenamefont {Friesem}, \citenamefont {Raz},\ and\
		\citenamefont {Davidson}}]{Gershenzon:2020}%
	\BibitemOpen
	\bibfield  {author} {\bibinfo {author} {\bibfnamefont {I.}~\bibnamefont
			{Gershenzon}}, \bibinfo {author} {\bibfnamefont {G.}~\bibnamefont {Arwas}},
		\bibinfo {author} {\bibfnamefont {S.}~\bibnamefont {Gadasi}}, \bibinfo
		{author} {\bibfnamefont {C.}~\bibnamefont {Tradonsky}}, \bibinfo {author}
		{\bibfnamefont {A.}~\bibnamefont {Friesem}}, \bibinfo {author} {\bibfnamefont
			{O.}~\bibnamefont {Raz}},\ and\ \bibinfo {author} {\bibfnamefont
			{N.}~\bibnamefont {Davidson}},\ }\bibfield  {title} {\bibinfo {title} {Exact
			mapping between a laser network loss rate and the classical {XY}
			{H}amiltonian by laser loss control},\ }\href
	{https://doi.org/doi:10.1515/nanoph-2020-0137} {\bibfield  {journal}
		{\bibinfo  {journal} {Nanophotonics}\ }\textbf {\bibinfo {volume} {9}},\
		\bibinfo {pages} {4117} (\bibinfo {year} {2020})}\BibitemShut {NoStop}%
	\bibitem [{\citenamefont {Kurtscheid}\ \emph {et~al.}(2019)\citenamefont
		{Kurtscheid}, \citenamefont {Dung}, \citenamefont {Busley}, \citenamefont
		{Vewinger}, \citenamefont {Rosch},\ and\ \citenamefont
		{Weitz}}]{Kurtscheid:2019}%
	\BibitemOpen
	\bibfield  {author} {\bibinfo {author} {\bibfnamefont {C.}~\bibnamefont
			{Kurtscheid}}, \bibinfo {author} {\bibfnamefont {D.}~\bibnamefont {Dung}},
		\bibinfo {author} {\bibfnamefont {E.}~\bibnamefont {Busley}}, \bibinfo
		{author} {\bibfnamefont {F.}~\bibnamefont {Vewinger}}, \bibinfo {author}
		{\bibfnamefont {A.}~\bibnamefont {Rosch}},\ and\ \bibinfo {author}
		{\bibfnamefont {M.}~\bibnamefont {Weitz}},\ }\bibfield  {title} {\bibinfo
		{title} {Thermally condensing photons into a coherently split state of
			light},\ }\href {https://doi.org/10.1126/science.aay1334} {\bibfield
		{journal} {\bibinfo  {journal} {Science}\ }\textbf {\bibinfo {volume}
			{366}},\ \bibinfo {pages} {894} (\bibinfo {year} {2019})}\BibitemShut
	{NoStop}%
	\bibitem [{\citenamefont {Jacqmin}\ \emph {et~al.}(2014)\citenamefont
		{Jacqmin}, \citenamefont {Carusotto}, \citenamefont {Sagnes}, \citenamefont
		{Abbarchi}, \citenamefont {Solnyshkov}, \citenamefont {Malpuech},
		\citenamefont {Galopin}, \citenamefont {Lema\^{\i}tre}, \citenamefont
		{Bloch},\ and\ \citenamefont {Amo}}]{Jacqmin:2014}%
	\BibitemOpen
	\bibfield  {author} {\bibinfo {author} {\bibfnamefont {T.}~\bibnamefont
			{Jacqmin}}, \bibinfo {author} {\bibfnamefont {I.}~\bibnamefont {Carusotto}},
		\bibinfo {author} {\bibfnamefont {I.}~\bibnamefont {Sagnes}}, \bibinfo
		{author} {\bibfnamefont {M.}~\bibnamefont {Abbarchi}}, \bibinfo {author}
		{\bibfnamefont {D.~D.}\ \bibnamefont {Solnyshkov}}, \bibinfo {author}
		{\bibfnamefont {G.}~\bibnamefont {Malpuech}}, \bibinfo {author}
		{\bibfnamefont {E.}~\bibnamefont {Galopin}}, \bibinfo {author} {\bibfnamefont
			{A.}~\bibnamefont {Lema\^{\i}tre}}, \bibinfo {author} {\bibfnamefont
			{J.}~\bibnamefont {Bloch}},\ and\ \bibinfo {author} {\bibfnamefont
			{A.}~\bibnamefont {Amo}},\ }\bibfield  {title} {\bibinfo {title} {Direct
			observation of dirac cones and a flatband in a honeycomb lattice for
			polaritons},\ }\href {https://doi.org/10.1103/PhysRevLett.112.116402}
	{\bibfield  {journal} {\bibinfo  {journal} {Phys. Rev. Lett.}\ }\textbf
		{\bibinfo {volume} {112}},\ \bibinfo {pages} {116402} (\bibinfo {year}
		{2014})}\BibitemShut {NoStop}%
	\bibitem [{\citenamefont {Vretenar}\ \emph
		{et~al.}(2021{\natexlab{b}})\citenamefont {Vretenar}, \citenamefont
		{Kassenberg}, \citenamefont {Bissesar}, \citenamefont {Toebes},\ and\
		\citenamefont {Klaers}}]{Vretenar:2021b}%
	\BibitemOpen
	\bibfield  {author} {\bibinfo {author} {\bibfnamefont {M.}~\bibnamefont
			{Vretenar}}, \bibinfo {author} {\bibfnamefont {B.}~\bibnamefont
			{Kassenberg}}, \bibinfo {author} {\bibfnamefont {S.}~\bibnamefont
			{Bissesar}}, \bibinfo {author} {\bibfnamefont {C.}~\bibnamefont {Toebes}},\
		and\ \bibinfo {author} {\bibfnamefont {J.}~\bibnamefont {Klaers}},\
	}\bibfield  {title} {\bibinfo {title} {Controllable josephson junction for
			photon {B}ose-{E}instein condensates},\ }\href
	{https://doi.org/10.1103/PhysRevResearch.3.023167} {\bibfield  {journal}
		{\bibinfo  {journal} {Phys. Rev. Res.}\ }\textbf {\bibinfo {volume} {3}},\
		\bibinfo {pages} {023167} (\bibinfo {year} {2021}{\natexlab{b}})}\BibitemShut
	{NoStop}%
	\bibitem [{\citenamefont {Lagoudakis}\ and\ \citenamefont
		{Berloff}(2017)}]{Lagoudakis:2017}%
	\BibitemOpen
	\bibfield  {author} {\bibinfo {author} {\bibfnamefont {P.~G.}\ \bibnamefont
			{Lagoudakis}}\ and\ \bibinfo {author} {\bibfnamefont {N.~G.}\ \bibnamefont
			{Berloff}},\ }\bibfield  {title} {\bibinfo {title} {A polariton graph
			simulator},\ }\href {https://doi.org/10.1088/1367-2630/aa924b} {\bibfield
		{journal} {\bibinfo  {journal} {New J. Phys.}\ }\textbf {\bibinfo {volume}
			{19}},\ \bibinfo {pages} {125008} (\bibinfo {year} {2017})}\BibitemShut
	{NoStop}%
	\bibitem [{\citenamefont {Mukherjee}\ \emph {et~al.}(2019)\citenamefont
		{Mukherjee}, \citenamefont {Myers}, \citenamefont {Lena}, \citenamefont
		{Ozden}, \citenamefont {Beaumariage}, \citenamefont {Sun}, \citenamefont
		{Steger}, \citenamefont {Pfeiffer}, \citenamefont {West}, \citenamefont
		{Daley},\ and\ \citenamefont {Snoke}}]{Mukherjee:2019a}%
	\BibitemOpen
	\bibfield  {author} {\bibinfo {author} {\bibfnamefont {S.}~\bibnamefont
			{Mukherjee}}, \bibinfo {author} {\bibfnamefont {D.~M.}\ \bibnamefont
			{Myers}}, \bibinfo {author} {\bibfnamefont {R.~G.}\ \bibnamefont {Lena}},
		\bibinfo {author} {\bibfnamefont {B.}~\bibnamefont {Ozden}}, \bibinfo
		{author} {\bibfnamefont {J.}~\bibnamefont {Beaumariage}}, \bibinfo {author}
		{\bibfnamefont {Z.}~\bibnamefont {Sun}}, \bibinfo {author} {\bibfnamefont
			{M.}~\bibnamefont {Steger}}, \bibinfo {author} {\bibfnamefont {L.~N.}\
			\bibnamefont {Pfeiffer}}, \bibinfo {author} {\bibfnamefont {K.}~\bibnamefont
			{West}}, \bibinfo {author} {\bibfnamefont {A.~J.}\ \bibnamefont {Daley}},\
		and\ \bibinfo {author} {\bibfnamefont {D.~W.}\ \bibnamefont {Snoke}},\
	}\bibfield  {title} {\bibinfo {title} {Observation of nonequilibrium motion
			and equilibration in polariton rings},\ }\href
	{https://doi.org/10.1103/PhysRevB.100.245304} {\bibfield  {journal} {\bibinfo
			{journal} {Phys. Rev. B}\ }\textbf {\bibinfo {volume} {100}},\ \bibinfo
		{pages} {245304} (\bibinfo {year} {2019})}\BibitemShut {NoStop}%
	\bibitem [{\citenamefont {Mukherjee}\ \emph {et~al.}(2021)\citenamefont
		{Mukherjee}, \citenamefont {Kozin}, \citenamefont {Nalitov}, \citenamefont
		{Shelykh}, \citenamefont {Sun}, \citenamefont {Myers}, \citenamefont {Ozden},
		\citenamefont {Beaumariage}, \citenamefont {Steger}, \citenamefont
		{Pfeiffer}, \citenamefont {West},\ and\ \citenamefont
		{Snoke}}]{Mukherjee:2021}%
	\BibitemOpen
	\bibfield  {author} {\bibinfo {author} {\bibfnamefont {S.}~\bibnamefont
			{Mukherjee}}, \bibinfo {author} {\bibfnamefont {V.~K.}\ \bibnamefont
			{Kozin}}, \bibinfo {author} {\bibfnamefont {A.~V.}\ \bibnamefont {Nalitov}},
		\bibinfo {author} {\bibfnamefont {I.~A.}\ \bibnamefont {Shelykh}}, \bibinfo
		{author} {\bibfnamefont {Z.}~\bibnamefont {Sun}}, \bibinfo {author}
		{\bibfnamefont {D.~M.}\ \bibnamefont {Myers}}, \bibinfo {author}
		{\bibfnamefont {B.}~\bibnamefont {Ozden}}, \bibinfo {author} {\bibfnamefont
			{J.}~\bibnamefont {Beaumariage}}, \bibinfo {author} {\bibfnamefont
			{M.}~\bibnamefont {Steger}}, \bibinfo {author} {\bibfnamefont {L.~N.}\
			\bibnamefont {Pfeiffer}}, \bibinfo {author} {\bibfnamefont {K.}~\bibnamefont
			{West}},\ and\ \bibinfo {author} {\bibfnamefont {D.~W.}\ \bibnamefont
			{Snoke}},\ }\bibfield  {title} {\bibinfo {title} {Dynamics of spin
			polarization in tilted polariton rings},\ }\href
	{https://doi.org/10.1103/PhysRevB.103.165306} {\bibfield  {journal} {\bibinfo
			{journal} {Phys. Rev. B}\ }\textbf {\bibinfo {volume} {103}},\ \bibinfo
		{pages} {165306} (\bibinfo {year} {2021})}\BibitemShut {NoStop}%
	\bibitem [{\citenamefont {Wang}\ \emph {et~al.}(2021)\citenamefont {Wang},
		\citenamefont {Xu}, \citenamefont {Su}, \citenamefont {Peng}, \citenamefont
		{Wu}, \citenamefont {Liew},\ and\ \citenamefont {Xiong}}]{Wang:2021a}%
	\BibitemOpen
	\bibfield  {author} {\bibinfo {author} {\bibfnamefont {J.}~\bibnamefont
			{Wang}}, \bibinfo {author} {\bibfnamefont {H.}~\bibnamefont {Xu}}, \bibinfo
		{author} {\bibfnamefont {R.}~\bibnamefont {Su}}, \bibinfo {author}
		{\bibfnamefont {Y.}~\bibnamefont {Peng}}, \bibinfo {author} {\bibfnamefont
			{J.}~\bibnamefont {Wu}}, \bibinfo {author} {\bibfnamefont {T.~C.}\
			\bibnamefont {Liew}},\ and\ \bibinfo {author} {\bibfnamefont
			{Q.}~\bibnamefont {Xiong}},\ }\bibfield  {title} {\bibinfo {title}
		{Spontaneously coherent orbital coupling of counterrotating exciton
			polaritons in annular perovskite microcavities},\ }\href
	{https://doi.org/10.1038/s41377-021-00478-w} {\bibfield  {journal} {\bibinfo
			{journal} {Light: Science \& Applications}\ }\textbf {\bibinfo {volume}
			{10}},\ \bibinfo {pages} {45} (\bibinfo {year} {2021})}\BibitemShut {NoStop}%
	\bibitem [{\citenamefont {Kurtscheid}\ \emph {et~al.}(2020)\citenamefont
		{Kurtscheid}, \citenamefont {Dung}, \citenamefont {Redmann}, \citenamefont
		{Busley}, \citenamefont {Klaers}, \citenamefont {Vewinger}, \citenamefont
		{Schmitt},\ and\ \citenamefont {Weitz}}]{Kurtscheid:2020}%
	\BibitemOpen
	\bibfield  {author} {\bibinfo {author} {\bibfnamefont {C.}~\bibnamefont
			{Kurtscheid}}, \bibinfo {author} {\bibfnamefont {D.}~\bibnamefont {Dung}},
		\bibinfo {author} {\bibfnamefont {A.}~\bibnamefont {Redmann}}, \bibinfo
		{author} {\bibfnamefont {E.}~\bibnamefont {Busley}}, \bibinfo {author}
		{\bibfnamefont {J.}~\bibnamefont {Klaers}}, \bibinfo {author} {\bibfnamefont
			{F.}~\bibnamefont {Vewinger}}, \bibinfo {author} {\bibfnamefont
			{J.}~\bibnamefont {Schmitt}},\ and\ \bibinfo {author} {\bibfnamefont
			{M.}~\bibnamefont {Weitz}},\ }\bibfield  {title} {\bibinfo {title} {Realizing
			arbitrary trapping potentials for light via direct laser writing of mirror
			surface profiles},\ }\href {https://doi.org/10.1209/0295-5075/130/54001}
	{\bibfield  {journal} {\bibinfo  {journal} {{EPL} (Europhys. Lett.)}\
		}\textbf {\bibinfo {volume} {130}},\ \bibinfo {pages} {54001} (\bibinfo
		{year} {2020})}\BibitemShut {NoStop}%
	\bibitem [{\citenamefont {Kirton}\ and\ \citenamefont
		{Keeling}(2015)}]{Kirton:2015}%
	\BibitemOpen
	\bibfield  {author} {\bibinfo {author} {\bibfnamefont {P.}~\bibnamefont
			{Kirton}}\ and\ \bibinfo {author} {\bibfnamefont {J.}~\bibnamefont
			{Keeling}},\ }\bibfield  {title} {\bibinfo {title} {Thermalization and
			breakdown of thermalization in photon condensates},\ }\href
	{https://doi.org/10.1103/PhysRevA.91.033826} {\bibfield  {journal} {\bibinfo
			{journal} {Phys. Rev. A}\ }\textbf {\bibinfo {volume} {91}},\ \bibinfo
		{pages} {033826} (\bibinfo {year} {2015})}\BibitemShut {NoStop}%
	\bibitem [{\citenamefont {Schmitt}\ \emph {et~al.}(2015)\citenamefont
		{Schmitt}, \citenamefont {Damm}, \citenamefont {Dung}, \citenamefont
		{Vewinger}, \citenamefont {Klaers},\ and\ \citenamefont
		{Weitz}}]{Schmitt:2015}%
	\BibitemOpen
	\bibfield  {author} {\bibinfo {author} {\bibfnamefont {J.}~\bibnamefont
			{Schmitt}}, \bibinfo {author} {\bibfnamefont {T.}~\bibnamefont {Damm}},
		\bibinfo {author} {\bibfnamefont {D.}~\bibnamefont {Dung}}, \bibinfo {author}
		{\bibfnamefont {F.}~\bibnamefont {Vewinger}}, \bibinfo {author}
		{\bibfnamefont {J.}~\bibnamefont {Klaers}},\ and\ \bibinfo {author}
		{\bibfnamefont {M.}~\bibnamefont {Weitz}},\ }\bibfield  {title} {\bibinfo
		{title} {Thermalization kinetics of light: From laser dynamics to equilibrium
			condensation of photons},\ }\href
	{https://doi.org/10.1103/PhysRevA.92.011602} {\bibfield  {journal} {\bibinfo
			{journal} {Phys. Rev. A}\ }\textbf {\bibinfo {volume} {92}},\ \bibinfo
		{pages} {011602} (\bibinfo {year} {2015})}\BibitemShut {NoStop}%
	\bibitem [{Sup()}]{Supplementary}%
	\BibitemOpen
	\href@noop {} {\bibinfo {title} {See supplemental material for details on
			experimental and theoretical methods, which includes refs.
			\cite{Stepanov:1971, Lakowicz:2006, Yokoyama:1989, Klaers:2011, Kirton:2013,
				Gioia:1997, Greveling:2017, Moodie:2017}}}\BibitemShut {NoStop}%
	\bibitem [{\citenamefont {Stepanov}\ and\ \citenamefont
		{Kazachenko}(1971)}]{Stepanov:1971}%
	\BibitemOpen
	\bibfield  {author} {\bibinfo {author} {\bibfnamefont {B.~I.}\ \bibnamefont
			{Stepanov}}\ and\ \bibinfo {author} {\bibfnamefont {L.~P.}\ \bibnamefont
			{Kazachenko}},\ }\bibfield  {title} {\bibinfo {title} {Universal relationship
			between absorption and emission spectra taking the solvent effect into
			account},\ }\href {https://doi.org/10.1007/BF00605796} {\bibfield  {journal}
		{\bibinfo  {journal} {Journal of Applied Spectroscopy}\ }\textbf {\bibinfo
			{volume} {14}},\ \bibinfo {pages} {596} (\bibinfo {year} {1971})}\BibitemShut
	{NoStop}%
	\bibitem [{\citenamefont {Lakowicz}(2006)}]{Lakowicz:2006}%
	\BibitemOpen
	\bibfield  {author} {\bibinfo {author} {\bibfnamefont {J.}~\bibnamefont
			{Lakowicz}},\ }\href {https://doi.org/10.1007/978-0-387-46312-4} {\emph
		{\bibinfo {title} {Principles of Fluorescence Spectroscopy}}}\ (\bibinfo
	{publisher} {Springer New York, NY},\ \bibinfo {year} {2006})\BibitemShut
	{NoStop}%
	\bibitem [{\citenamefont {Yokoyama}\ and\ \citenamefont
		{Brorson}(1989)}]{Yokoyama:1989}%
	\BibitemOpen
	\bibfield  {author} {\bibinfo {author} {\bibfnamefont {H.}~\bibnamefont
			{Yokoyama}}\ and\ \bibinfo {author} {\bibfnamefont {S.~D.}\ \bibnamefont
			{Brorson}},\ }\bibfield  {title} {\bibinfo {title} {{Rate equation analysis
				of microcavity lasers}},\ }\href {https://doi.org/10.1063/1.343793}
	{\bibfield  {journal} {\bibinfo  {journal} {Journal of Applied Physics}\
		}\textbf {\bibinfo {volume} {66}},\ \bibinfo {pages} {4801} (\bibinfo {year}
		{1989})}\BibitemShut {NoStop}%
	\bibitem [{\citenamefont {Klaers}\ \emph {et~al.}(2011)\citenamefont {Klaers},
		\citenamefont {Schmitt}, \citenamefont {Damm}, \citenamefont {Vewinger},\
		and\ \citenamefont {Weitz}}]{Klaers:2011}%
	\BibitemOpen
	\bibfield  {author} {\bibinfo {author} {\bibfnamefont {J.}~\bibnamefont
			{Klaers}}, \bibinfo {author} {\bibfnamefont {J.}~\bibnamefont {Schmitt}},
		\bibinfo {author} {\bibfnamefont {T.}~\bibnamefont {Damm}}, \bibinfo {author}
		{\bibfnamefont {F.}~\bibnamefont {Vewinger}},\ and\ \bibinfo {author}
		{\bibfnamefont {M.}~\bibnamefont {Weitz}},\ }\bibfield  {title} {\bibinfo
		{title} {Bose-{E}instein condensation of paraxial light},\ }\href
	{https://doi.org/10.1007/s00340-011-4734-6} {\bibfield  {journal} {\bibinfo
			{journal} {Appl. Phys. B: Lasers Opt.}\ }\textbf {\bibinfo {volume} {105}},\
		\bibinfo {pages} {17} (\bibinfo {year} {2011})}\BibitemShut {NoStop}%
	\bibitem [{\citenamefont {Kirton}\ and\ \citenamefont
		{Keeling}(2013)}]{Kirton:2013}%
	\BibitemOpen
	\bibfield  {author} {\bibinfo {author} {\bibfnamefont {P.}~\bibnamefont
			{Kirton}}\ and\ \bibinfo {author} {\bibfnamefont {J.}~\bibnamefont
			{Keeling}},\ }\bibfield  {title} {\bibinfo {title} {Nonequilibrium model of
			photon condensation},\ }\href
	{https://doi.org/10.1103/PhysRevLett.111.100404} {\bibfield  {journal}
		{\bibinfo  {journal} {Phys. Rev. Lett.}\ }\textbf {\bibinfo {volume} {111}},\
		\bibinfo {pages} {100404} (\bibinfo {year} {2013})}\BibitemShut {NoStop}%
	\bibitem [{\citenamefont {Gioia}\ and\ \citenamefont
		{Ortiz}(1997)}]{Gioia:1997}%
	\BibitemOpen
	\bibfield  {author} {\bibinfo {author} {\bibfnamefont {G.}~\bibnamefont
			{Gioia}}\ and\ \bibinfo {author} {\bibfnamefont {M.}~\bibnamefont {Ortiz}},\
	}\bibfield  {title} {\bibinfo {title} {Delamination of compressed thin
			films}\ }(\bibinfo  {publisher} {Elsevier},\ \bibinfo {year} {1997})\ pp.\
	\bibinfo {pages} {119--192}\BibitemShut {NoStop}%
	\bibitem [{\citenamefont {Greveling}\ \emph {et~al.}(2017)\citenamefont
		{Greveling}, \citenamefont {van~der Laan}, \citenamefont {Jagers},\ and\
		\citenamefont {van Oosten}}]{Greveling:2017}%
	\BibitemOpen
	\bibfield  {author} {\bibinfo {author} {\bibfnamefont {S.}~\bibnamefont
			{Greveling}}, \bibinfo {author} {\bibfnamefont {F.}~\bibnamefont {van~der
				Laan}}, \bibinfo {author} {\bibfnamefont {H.~C.}\ \bibnamefont {Jagers}},\
		and\ \bibinfo {author} {\bibfnamefont {D.}~\bibnamefont {van Oosten}},\
	}\href@noop {} {\bibinfo {title} {Polarization of a bose-einstein condensate
			of photons in a dye-filled microcavity}} (\bibinfo {year} {2017}),\ \Eprint
	{https://arxiv.org/abs/1712.08426} {arXiv:1712.08426 [cond-mat.quant-gas]}
	\BibitemShut {NoStop}%
	\bibitem [{\citenamefont {Moodie}\ \emph {et~al.}(2017)\citenamefont {Moodie},
		\citenamefont {Kirton},\ and\ \citenamefont {Keeling}}]{Moodie:2017}%
	\BibitemOpen
	\bibfield  {author} {\bibinfo {author} {\bibfnamefont {R.~I.}\ \bibnamefont
			{Moodie}}, \bibinfo {author} {\bibfnamefont {P.}~\bibnamefont {Kirton}},\
		and\ \bibinfo {author} {\bibfnamefont {J.}~\bibnamefont {Keeling}},\
	}\bibfield  {title} {\bibinfo {title} {Polarization dynamics in a photon
			bose-einstein condensate},\ }\href
	{https://doi.org/10.1103/PhysRevA.96.043844} {\bibfield  {journal} {\bibinfo
			{journal} {Phys. Rev. A}\ }\textbf {\bibinfo {volume} {96}},\ \bibinfo
		{pages} {043844} (\bibinfo {year} {2017})}\BibitemShut {NoStop}%		
	\bibitem [{\citenamefont {Keeling}\ and\ \citenamefont
		{Kirton}(2016)}]{Keeling:2016}%
	\BibitemOpen
	\bibfield  {author} {\bibinfo {author} {\bibfnamefont {J.}~\bibnamefont
			{Keeling}}\ and\ \bibinfo {author} {\bibfnamefont {P.}~\bibnamefont
			{Kirton}},\ }\bibfield  {title} {\bibinfo {title} {Spatial dynamics,
			thermalization, and gain clamping in a photon condensate},\ }\href
	{https://doi.org/10.1103/PhysRevA.93.013829} {\bibfield  {journal} {\bibinfo
			{journal} {Phys. Rev. A}\ }\textbf {\bibinfo {volume} {93}},\ \bibinfo
		{pages} {013829} (\bibinfo {year} {2016})}\BibitemShut {NoStop}%
	\bibitem [{\citenamefont {Majumdar}\ and\ \citenamefont
		{Gerace}(2013)}]{Majumdar:2013}%
	\BibitemOpen
	\bibfield  {author} {\bibinfo {author} {\bibfnamefont {A.}~\bibnamefont
			{Majumdar}}\ and\ \bibinfo {author} {\bibfnamefont {D.}~\bibnamefont
			{Gerace}},\ }\bibfield  {title} {\bibinfo {title} {Single-photon blockade in
			doubly resonant nanocavities with second-order nonlinearity},\ }\href
	{https://doi.org/10.1103/PhysRevB.87.235319} {\bibfield  {journal} {\bibinfo
			{journal} {Phys. Rev. B}\ }\textbf {\bibinfo {volume} {87}},\ \bibinfo
		{pages} {235319} (\bibinfo {year} {2013})}\BibitemShut {NoStop}%
	\bibitem [{\citenamefont {Dung}\ \emph {et~al.}(2017)\citenamefont {Dung},
		\citenamefont {Kurtscheid}, \citenamefont {Damm}, \citenamefont {Schmitt},
		\citenamefont {Vewinger}, \citenamefont {Weitz},\ and\ \citenamefont
		{Klaers}}]{Dung:2017}%
	\BibitemOpen
	\bibfield  {author} {\bibinfo {author} {\bibfnamefont {D.}~\bibnamefont
			{Dung}}, \bibinfo {author} {\bibfnamefont {C.}~\bibnamefont {Kurtscheid}},
		\bibinfo {author} {\bibfnamefont {T.}~\bibnamefont {Damm}}, \bibinfo {author}
		{\bibfnamefont {J.}~\bibnamefont {Schmitt}}, \bibinfo {author} {\bibfnamefont
			{F.}~\bibnamefont {Vewinger}}, \bibinfo {author} {\bibfnamefont
			{M.}~\bibnamefont {Weitz}},\ and\ \bibinfo {author} {\bibfnamefont
			{J.}~\bibnamefont {Klaers}},\ }\bibfield  {title} {\bibinfo {title} {Variable
			potentials for thermalized light and coupled condensates},\ }\href
	{https://doi.org/10.1038/nphoton.2017.139} {\bibfield  {journal} {\bibinfo
			{journal} {Nat. Photon.}\ }\textbf {\bibinfo {volume} {11}},\ \bibinfo
		{pages} {565} (\bibinfo {year} {2017})}\BibitemShut {NoStop}%
	\bibitem [{\citenamefont {Busley}\ \emph {et~al.}(2022)\citenamefont {Busley},
		\citenamefont {Miranda}, \citenamefont {Redmann}, \citenamefont {Kurtscheid},
		\citenamefont {Umesh}, \citenamefont {Vewinger}, \citenamefont {Weitz},\ and\
		\citenamefont {Schmitt}}]{Busley:2022}%
	\BibitemOpen
	\bibfield  {author} {\bibinfo {author} {\bibfnamefont {E.}~\bibnamefont
			{Busley}}, \bibinfo {author} {\bibfnamefont {L.~E.}\ \bibnamefont {Miranda}},
		\bibinfo {author} {\bibfnamefont {A.}~\bibnamefont {Redmann}}, \bibinfo
		{author} {\bibfnamefont {C.}~\bibnamefont {Kurtscheid}}, \bibinfo {author}
		{\bibfnamefont {K.~K.}\ \bibnamefont {Umesh}}, \bibinfo {author}
		{\bibfnamefont {F.}~\bibnamefont {Vewinger}}, \bibinfo {author}
		{\bibfnamefont {M.}~\bibnamefont {Weitz}},\ and\ \bibinfo {author}
		{\bibfnamefont {J.}~\bibnamefont {Schmitt}},\ }\bibfield  {title} {\bibinfo
		{title} {Compressibility and the equation of state of an optical quantum gas
			in a box},\ }\href {https://doi.org/10.1126/science.abm2543} {\bibfield
		{journal} {\bibinfo  {journal} {Science}\ }\textbf {\bibinfo {volume}
			{375}},\ \bibinfo {pages} {1403} (\bibinfo {year} {2022})}\BibitemShut
	{NoStop}%
	\bibitem [{\citenamefont {Xue}\ \emph {et~al.}(2021)\citenamefont {Xue},
		\citenamefont {Chestnov}, \citenamefont {Sedov}, \citenamefont {Kiktenko},
		\citenamefont {Fedorov}, \citenamefont {Schumacher}, \citenamefont {Ma},\
		and\ \citenamefont {Kavokin}}]{Xue:2021}%
	\BibitemOpen
	\bibfield  {author} {\bibinfo {author} {\bibfnamefont {Y.}~\bibnamefont
			{Xue}}, \bibinfo {author} {\bibfnamefont {I.}~\bibnamefont {Chestnov}},
		\bibinfo {author} {\bibfnamefont {E.}~\bibnamefont {Sedov}}, \bibinfo
		{author} {\bibfnamefont {E.}~\bibnamefont {Kiktenko}}, \bibinfo {author}
		{\bibfnamefont {A.~K.}\ \bibnamefont {Fedorov}}, \bibinfo {author}
		{\bibfnamefont {S.}~\bibnamefont {Schumacher}}, \bibinfo {author}
		{\bibfnamefont {X.}~\bibnamefont {Ma}},\ and\ \bibinfo {author}
		{\bibfnamefont {A.}~\bibnamefont {Kavokin}},\ }\bibfield  {title} {\bibinfo
		{title} {Split-ring polariton condensates as macroscopic two-level quantum
			systems},\ }\href {https://doi.org/10.1103/PhysRevResearch.3.013099}
	{\bibfield  {journal} {\bibinfo  {journal} {Phys. Rev. Res.}\ }\textbf
		{\bibinfo {volume} {3}},\ \bibinfo {pages} {013099} (\bibinfo {year}
		{2021})}\BibitemShut {NoStop}%
	\bibitem [{\citenamefont {Fang}\ \emph {et~al.}(2012)\citenamefont {Fang},
		\citenamefont {Yu},\ and\ \citenamefont {Fan}}]{Fang:2012}%
	\BibitemOpen
	\bibfield  {author} {\bibinfo {author} {\bibfnamefont {K.}~\bibnamefont
			{Fang}}, \bibinfo {author} {\bibfnamefont {Z.}~\bibnamefont {Yu}},\ and\
		\bibinfo {author} {\bibfnamefont {S.}~\bibnamefont {Fan}},\ }\bibfield
	{title} {\bibinfo {title} {Realizing effective magnetic field for photons by
			controlling the phase of dynamic modulation},\ }\href
	{https://doi.org/10.1038/nphoton.2012.236} {\bibfield  {journal} {\bibinfo
			{journal} {Nat. Photonics}\ }\textbf {\bibinfo {volume} {6}},\ \bibinfo
		{pages} {782} (\bibinfo {year} {2012})}\BibitemShut {NoStop}%
	\bibitem [{\citenamefont {Gladilin}\ and\ \citenamefont
		{Wouters}(2020)}]{Gladilin:2020b}%
	\BibitemOpen
	\bibfield  {author} {\bibinfo {author} {\bibfnamefont {V.~N.}\ \bibnamefont
			{Gladilin}}\ and\ \bibinfo {author} {\bibfnamefont {M.}~\bibnamefont
			{Wouters}},\ }\bibfield  {title} {\bibinfo {title} {Vortices in
			nonequilibrium photon condensates},\ }\href
	{https://doi.org/10.1103/PhysRevLett.125.215301} {\bibfield  {journal}
		{\bibinfo  {journal} {Phys. Rev. Lett.}\ }\textbf {\bibinfo {volume} {125}},\
		\bibinfo {pages} {215301} (\bibinfo {year} {2020})}\BibitemShut {NoStop}%
\end{thebibliography}
\end{document}